\documentclass[11pt,reqno]{amsart}

\usepackage{graphicx}% Include figure files
\usepackage{bm}% bold math

\usepackage{color}
\usepackage{tikz-cd}
\usepackage{amsfonts}
\usepackage{amsmath}
\usepackage{amssymb}
\usepackage{graphicx}
\usepackage{bm}
\usepackage{latexsym}
\usepackage{bbm,dsfont}
\usepackage{amscd}
\usepackage{stmaryrd}
\usepackage{tikz}
\usepackage[arrow,matrix,curve,cmtip,ps]{xy}
\usepackage{pb-diagram} 
\usepackage{pb-xy}

\newtheorem{proposition}{Proposition}
\newtheorem{theorem}{Theorem}

\newtheorem{corollary}{Corollary}
\newtheorem{definition}{Definition}

\newcommand{\ip[2]}{\langle{#1}|{#2}\rangle}

\newcommand{\state}[1]{\left\vert #1 \right\rangle}
\newcommand{\dstate}[1]{\left\langle #1 \right\vert}

\newcommand{\id}{\mathbbm{1}}

\begin{document}

%\preprint{APS/123-QED}

\title{Topos logic in measurement-based quantum computation}% Force line breaks with \\

\maketitle

\begin{center}\author
{
Leon Loveridge$^{1,2}$,
Raouf Dridi$^{2}$,
Robert Raussendorf$^{2}$}
\end{center}
\bigskip
\address{
$^{1}$Department of Computer Science, University of Oxford, Oxford, UK.\\
$^{2}$ Department of Physics and Astronomy, University of British Columbia, Vancouver, Canada.}

\date{\today}% It is always \today, today,

\begin{abstract}
We report first steps towards elucidating the relationship between contextuality, measurement-based quantum computation (MBQC) and
the non-classical logic of a topos associated with the computation. We show that, in a class of MBQC, classical universality \emph{requires} non-classical logic, which is ``consumed" during the course of the computation, thereby pinpointing
another potential quantum computational resource.
\end{abstract}
%%%%%%%%%%%%%%%%%%%%%%%%%%%

%%%%%%%%%% Insert the texts which can accomdate on firstpage in the tag "fmtext" %%%%%

\section{Introduction}
%%%% Insert A head here
The quantum mechanical description of the world has  startling features which are quite alien from a classical
point of view. Epitomising the structural differences from classical physics includes the existence of entangled states, non-commutative observable algebras and the necessary disturbances caused by measurement. Less well known is the impossibility of assigning values to quantum observables whilst preserving some basic functional relations -- a trait known as \emph{contextuality} and a central concept in this paper.

An upshot of the mathematical structure of quantum theory is the possibility of attaining certain advantages over classical physics in regard to computation; there now exist numerous examples of quantum protocols which far outperform their classical
analogues, which are well documented in the literature.  
However, the origin of the quantum advantage remains unclear; ``largeness" of Hilbert space, superposition, interference,
and entanglement have been suggested as candidates in the past. Whilst being supported in a variety of situations 
\cite{cemm, gvid1, vwd3}, each of these has also met with objection \cite{Gphd, bmw1, gfe1, mvdn1}.
Recently, \emph{contextuality} has been proposed as a
contender for the source of the speedup in quantum computation; it has been observed that for quantum computation with a
restricted gate set and injection of magic states \cite{bk1}, contextuality
is necessary for both quantum computational universality and the hardness
of classical simulation of quantum computation \cite{vwfe1, hwve1}.

Contextuality also plays a role in measurement-based quantum computation
(MBQC). If an MBQC on qubits 
evaluates a non-linear Boolean function (in which case the computation is classically universal)
with high probability of success, this computation is also contextual
\cite{rau1}. Therefore, for MBQC, contextuality is important not only for what can be computed efficiently,
but for what can be computed at all. The simplest example of this is the execution of an OR-gate using
Pauli measurements on a 3-qubit Greenberger-Horne-Zeilinger (GHZ) state \cite{ab1}.
This gate is Mermin's proof \cite{Merm} of the Kochen-Specker theorem in Hilbert space-dimension 8, recast as a simple quantum computation.

In this paper we offer a perspective related to contextuality but distinct from it: that a quantum-over-classical 
advantage arises from the \emph{non-classical internal logic} of the given computation (we shall treat ``classical" and ``Boolean" as synonyms). Specifically, we associate a particular topos to the computation which comes equipped
with a natural logical structure, and show that the topos logic is non-Boolean whenever contextuality is present.
With \cite{rau1} we find that if the topos logic is Boolean, the MBQC may only compute linear functions. Hence non-Booleanness is a \emph{necessary condition} for quantum computational advantage in the class of computations we discuss. We also
demonstrate explicitly how the non-Booleanness is depleted by the act of computation, endorsing the view that non-Booleanness
is a quantum computational resource.

Boolean logic was devised in the first instance
as a formalisation of the ``laws of Thought" \cite{boo1}, i.e., as a way of consistently reasoning about the (classical) world. Many years have passed since Boole's seminal contribution;
a modern connection to Boolean logic is the theory of (classical, digital) computation. The quantum analogue---the relationship between quantum computation and ``quantum logic"---has thus far remained largely unexplored (though there are exceptions; see e.g.
\cite{gre1} for a version of ``quantum computational logic"). In this paper we make some inroads into this territory. 

Soon after the mathematical structure of quantum mechanics was essentially crystallised it was realised that the logical structure of the theory was fundamentally different from that found in classical physics \cite{bvn1}.
Though there are now many candidates for the ``logic of quantum theory" (see, e.g.,\cite{gre1}), we choose to follow the topos approach initiated by
Isham and Butterfield and developed by others (see, e.g., \cite{id1}, \cite{Flo1}; there are numerous other topos approaches - Adelman and Corbett \cite{cor1} seem to have been the first on the scene of ``topos quantum mechanics", and the work
of  Heunen, Landsman, and Spitters \cite{hls1} takes a covariant, rather than presheaf, approach). Our reasons for this choice are manifold. The logic of a topos is distributive, in contrast to the (propositional) logic ingeniously fashioned from projections by Birkhoff and von Neumann \cite{bvn1},  which seems to have permanently acquired the name ``quantum logic". 

Moreover, topos quantum logic is a first order predicate logic, thus reaching well beyond the propositional quantum logic and allowing for the expression of a much larger class of statements (involving quantifiers, for instance). It is in the topos approach that the connection between logic and contextuality is most clearly seen; we have a simple argument demonstrating that wherever there is contextuality there is non-Boolean topos logic, and that this applies to both state-independent and state-dependent versions of contextuality. Furthermore, for the topoi of interest in this investigation, the quantification of non-Booleanness takes a simple form, allowing for a direct demonstration in the Anders-Browne OR-gate that the ``amount" of non-Booleanness present is related to computational power at each stage of the computation.

\section{Survey of Results}
Before proceeding to the provision of the mathematical and physical frameworks within which our work is carried out,
we believe it helpful to survey the main results of this paper, in part as useful reference material and summary, and in part as a guide for what follows.

Following the pioneering work of Isham, Butterfield and D\"{o}ring, we begin by discussing the Kochen-Specker theorem
in terms of the ``spectral presheaf" $\Sigma$, which associates to each Abelian observable algebra (or \emph{context}) a classical phase space, 
viewed as the space of \emph{valuations} of the given observables. The Kochen-Specker theorem provides a negative answer to the question ``is it possible that observables in quantum mechanics have predetermined measurement outcomes which are preserved under functional relations between these observables?". The statement
of the Kochen-Specker theorem can be recast in terms of the (non-)existence of global sections of $\Sigma$. We are also
able, through the \emph{pseudostate} $\mathfrak{w}^{\state{\varphi}}$, to give state-dependent Kochen-Specker proofs from within the presheaf perspective, providing a simple version of a proof of Mermin in the new language. We note that what is required
for such a proof is a specific subposet $\mathcal{W(H)}$ of the poset $\mathcal{V(H)}$ of all Abelian subalgebras of $B(\mathcal{H})$ - the space of all linear operators on a finite dimensional Hilbert space $\mathcal{H}$. $\mathcal{W(H)}$ is also related to computation, constructed by taking intersections of lists of local observables whose product has the computational resource state as an eigenstate, and then forming the Abelian algebras at each node via the double commutant.

Understanding how $\Sigma$ and $\mathfrak{w}^{\state{\varphi}}$ relate to their ``environment" amounts to constructing the \emph{topos} $\widehat{\mathcal{W(H)}}$ of all set-valued presheaves on $\mathcal{W(H)}$. A topos is a set-like universe; from a logical perspective any topos behaves similarly to the category 
of sets but fails in two crucial respects: there are many possible truth values, arranged in a so-called \emph{Heyting algebra}, and the law of excluded middle fails. The latter
is equivalent to the statement that the internal logic of the given topos is \emph{non-Boolean}. Moreover, the axiom of choice (AC) fails in most topoi and, by a theorem of Diaconescu, choice and excluded middle are related; in any topos choice implies excluded middle and hence Booleanness. In the topoi of interest to us, namely the topoi of presheaves on a poset, 
the above implication is promoted to an equivalence: Boolean logic is equivalent to AC. By considering the appropriate categorical formulation of choice, we arrive at our first main observation: \emph{the presence of state-dependent or state-independent contextuality implies non-Boolean logic in $\widehat{\mathcal{W(H)}}$}.

Through the intermediary of contextuality, the relationship between MBQC and logic is considered. 
Anders and Browne \cite{ab1} noticed that the state-dependent Kochen-Specker proof of Mermin can be recast as an MBQC
OR-gate, for which they observe that contextuality is a necessary requirement. For a general $\ell$-2 MBQC (see section \ref{sec:mbqc}), it has been shown \cite{rau1} that non-linear function evaluation implies contextuality in the given computation, i.e., the existence
of a Kochen-Specker proof. By constructing the poset $\mathcal{W(H)}$ canonically associated to the MBQC and identifying the corresponding topos,
we may call upon the results stated in the previous paragraph to demonstrate that:\\

\emph{For the computation of non-linear
functions in $\ell$-2 MBQC, the associated topos is necessarily non-Boolean.}\\

 In other words, just as entanglement, superposition and contextuality may be understood to be computational resources, we see that non-Boolean topos logic may be viewed similarly.
By considering each stage of the computation  in the Anders-Browne example, and calling upon a construction which quantifies
the non-Booleanness of the associated topos at each stage, we show that the non-Booleanness is 
depleted by the computation and is directly related to the class of functions which can be computed after each measurement. Many of our findings are generic in the class of $\ell$-2 MBQC we consider.\\

The rest of this paper is organised as follows. In section \ref{cs} we give definitions of state-independent and state-dependent contextuality in terms of $\Sigma$ and 
$\mathfrak{w}^{\state{\varphi}}$, respectively, including Mermin's state-dependent proof from this perspective. Section
\ref{sec:mbqc} provides the definition of $\ell$-2 MBQC as well as Anders and Browne's contextual OR-gate. We then turn to
topoi and logic in section \ref{sec:tal}. Motivating the notion of a Heyting algebra from the propositional structure governed
by the open sets of a topological space, we then provide details of the topos $\mathsf{Sets}$ required for understanding
the logical structures of a general topos which will be important in later sections. We provide some important definitions
in the general theory of topoi and give the topos version of AC.

A consideration of the topos $\widehat{P}$ of presheaves on a generic poset $P$ and its quantum realisation $\mathcal{W(H)}$ constitutes section \ref{sec:prspos}. 
We present an important isomorphism between $\widehat{P}$ and the topos of sheaves $Sh(\mathsf{P})$ on a topological space $\mathsf{P}$ constructed from the poset $P$, which allows for a simple quantification of the non-Booleanness of $\widehat{P}$.
Section \ref{sec:prspos} also establishes the necessity of non-Boolean logic for non-linear function evaluation in $\ell$-2 MBQC,
and section \ref{sec:logcomp} demonstrates the consumption of the non-Booleanness during the execution of the computation.

Sections \ref{cs}, \ref{sec:mbqc} and \ref{sec:tal} contain review material on contextuality, MBQC and topoi respectively as one or more topic may be unfamiliar. We include a summary paragraph at the beginning to highlight the essential content.

\section{Contextuality in terms of $\Sigma$ and $\mathfrak{w}^{\state{\varphi}}$}\label{cs}

In this section we formulate the state-independent and state-dependent Kochen-Specker theorems from the perspective of the Isham-Butterfield
presheaves $\Sigma$ and $\mathfrak{w}^{\state{\varphi}}$, and demonstrate that a simple proof of Mermin \cite{Merm} may be cast in this language. The main observation is that only a small subposet of the poset of all Abelian subalgebras of $B({\mathcal{H})}$ is required to demonstrate a contradiction with the assumption of pre-assigned measurement outcomes.

We work in a complex Hilbert space $\mathcal{H}$ for which $\dim \mathcal{H} >2$, equipped with the standard
inner product $\ip[\cdot]{\cdot}: \mathcal{H}\times \mathcal{H} \to \mathbb{C}$, linear in the second argument,
pure states being given as unit vectors in $\mathcal{H}$. In order to obviate various technical topological issues we also assume that $\dim \mathcal{H}< \infty$. This is also justified on physical grounds; the quantum computational protocols we consider all take place in the finite dimensional setting.

 Let $B(\mathcal{H})$ denote the algebra of all linear mappings $\mathcal{H} \to \mathcal{H}$, and $\mathcal{V}(\mathcal{H})$ the collection of all unital \emph{Abelian} subalgebras of $B(\mathcal{H})$, (partially) ordered by inclusion. To each $V \in \mathcal{V(H)}$ ($V$ viewed as a $C^*$ algebra)
we form the set of \emph{characters} $\Sigma (V)$;
\begin{equation}
\Sigma (V):=\{\lambda_V: V \to \mathbb{C}~|~\lambda_V \neq 0,~ \lambda_V~ \text{is linear and } \lambda_V(ab)=\lambda_V(a)\lambda_V(b)\}.
\end{equation}
Following Isham {\it et al.} (e.g., \cite{id1}) we also define how $\Sigma$ acts on inclusion relations between different Abelian subalgebras: for $\lambda_V \in \Sigma (V)$, if
$U \subseteq V$, define 
\begin{align}
&\Sigma (U\subseteq V) (\lambda_V):=\lambda_V|_U~ \text{(as function restriction)}\\
&\lambda_V \mapsto \lambda_V|_U;
\end{align}
the mapping $\Sigma (U \subseteq V)$ is called a \emph{restriction map}. It is clear that, if $U \subseteq V \subseteq W$ (all Abelian), 
\begin{equation}\label{eq:dr}
\Sigma (V \subseteq W) \circ \Sigma (U \subseteq V) = \Sigma (U \subseteq W);
\end{equation} in other words, it makes no difference if one restricts in
stages, or ``all at once". In mathematical terms, 
 $\Sigma$ is then referred to as a \emph{contravariant, set-valued functor} (or \emph{presheaf}), called the \emph{spectral presheaf} and represents the ``state space", as introduced in \cite{ib3}.

For each $V \in \mathcal{V(H)}$, $\Sigma (V)$ can be viewed both topologically and logically as a classical phase space. 
It is topologically compact and Hausdorff, and the algebra of observables $V$ can be recovered as the space
of continuous functions $\Sigma (V) \to \mathbb{C}$ (in other words, $V \cong C(\Sigma (V)) $ and the isomorphism is isometric; this is the \emph{Gelfand duality} arising from the \emph{Gelfand transform} $A \mapsto \bar {A}$; $\lambda (A):=\bar{A}(\lambda)$). The algebra of physical propositions is given by the clopen (closed and open) subsets of $\Sigma (V)$ which is necessarily Boolean, reflecting the standard propositional structure corresponding to classical physical systems. 

\subsection{Valuations and Sections}
The set $\Sigma (V)$---the \emph{Gelfand spectrum} of $V$---is in bijection with the set of simultaneous eigenvectors of self-adjoint operators in $V$,
and therefore, for any self-adjoint $a \in V$ and any $\lambda_V \in \Sigma (V)$, $\lambda_V (a)$ is an eigenvalue of $a$ (we call this condition \emph{SPEC}). Furthermore,
 any $\lambda_V$ obeys the functional constraint $f\left(\lambda_V(a)\right) = \lambda_V \left(f(a) \right)$ for any (continuous) function $f:\mathbb{R} \to \mathbb{R}$ and
self-adjoint $a$ in $V$. Hence $\lambda_V$ acting on $V$ satisfies the ``functional composition principle" \emph{FUNC} (see \cite{red1}, ch. 5, for details) and can therefore be viewed as a \emph{valuation} of the self-adjoint operators in
$V$ (and the collection of all $\lambda_V$ are all the valuations). Such valuations are called \emph{local sections} (above $V$). FUNC states that  if an observable $a$ (energy, for example) has value $a_0$, then $f(a)$ has value $f(a_0)$ (i.e., the value of the square of the energy should be the square of its value, for instance), and is  (along with the spectrum requirement SPEC) an ostensibly mild condition even when considered on the whole of $B(\mathcal{H})$. However, $\mathcal{V(H)}$  is woven together in a complicated way (recall the partial order) and, as we shall see, this prohibits the existence of ``global" valuations, i.e., those defined on all observables in $B(\mathcal{H})$.

The existence of a valuation on $B(\mathcal{H})$  is of clear importance
in attempts to restore any kind of realism to quantum theory: if observables can be said to have values prior to and independently of measurement,
the status of measurement can be downgraded to mere revelation of values/properties ``held" by the quantum object in question.
If such a valuation were to exist, quantum theory would then retain some essence of classical physics where all observables have values in all states, and where propositions about a physical system are either true or false. To reiterate, a valuation $\lambda:B(\mathcal{H}) \to \mathbb{C}$
must satisfy:
\begin{itemize}
\item For $a=a^*$, $\lambda(a) \in \sigma (a)$ (i.e., the value must be an eigenvalue) (SPEC)
\item For $a=a^*$ and $f: \mathbb{R} \to \mathbb{R}$, $f\left (\lambda(a)\right )=\lambda \left( f(a)\right)$ (FUNC);
\end{itemize}
any such valuation is necessarily linear and multiplicative on commuting operators and corresponds to sections above Abelian algebras. It is automatic, of course, that for a projection $P$, $\lambda (P) \in \{0,1\}$.

If such a global value assignment $\lambda$ existed, we would be able to choose for each 
$V \in \mathcal{V(H)}$ a point $\lambda_V$ in $\Sigma (V)$ in such a way that, for any $U \subseteq V$, $\Sigma (U \subseteq V)(\lambda_V) = \lambda_U$. From equation \eqref{eq:dr}, we see that local valuations, on $U$, say, are not sensitive to 
which other Abelian algebras $U$ is contained in if they come from a global valuation, intuitively corresponding to 
a non-contextuality requirement.

\subsection{The Kochen-Specker theorem}

As we have seen, a global assignment $\lambda$ would specify an eigenvalue of each self-adjoint operator in $B(\mathcal{H})$ (SPEC)
whilst respecting functional relations (FUNC). Keeping the same notation we now view $\lambda$ as giving rise to a local section above each $V$, in such a way as to respect the inclusions in $\mathcal{V(H)}$.
\begin{definition}
A \emph{global section} $\lambda$ of $\Sigma$ is an assignment of a point $\lambda (V) \equiv \lambda_V \in \Sigma (V)$ for each $V \in \mathcal{V(H)}$, such that for $U \subseteq V$, $\Sigma (U \subseteq V)(\lambda_V) = \lambda_V|_U = \lambda_U$.
\end{definition}

There is a categorical description in which a global section looks like a ``point" of a set. Consider the singleton set
$\{ *\}$---the \emph{terminal object} in the category $\mathsf{Sets}$ of abstract sets and functions---characterised by the property that for any set $A$ (including the empty set $\emptyset$), there is
precisely one map $A \to \{ * \}$. Clearly any point of a set $A$ can be given as a map $\{ * \} \to A$, and the collection of all such maps determines $A$ completely. 

In the category of all presheaves $\mathcal{V(H)} \to \mathsf{Sets}=:\mathcal{E}$ (which is also a topos - more on this later), in order to define ``points" of objects we will need to also consider maps between presheaves, which are the arrows
in $\mathcal{E}$:
\begin{definition}
Let $\mathcal{F,G}: \mathcal{V(H)} \to \mathsf{Sets}$ be two presheaves. A natural transformation $\tau : F \to G$ is a map of presheaves,
given context-wise as $\tau_V :F(V) \to G(V)$, which for each $U \subseteq V$, $\tau_U \circ \mathcal{F}(U \subseteq V) = G(U \subseteq V) \circ \tau_V$.
\end{definition}
The constant presheaf $1 : \mathcal{V(H)} \to \mathsf{Sets}$ for which
$1(V)= \{*\}$ for each $V$ (and on inclusions $1(U \subseteq V) (\{*\}) = \{*\}$) is a terminal object in $\mathcal{E}$; for 
any presheaf $A$ in $\mathcal{E}$ there is precisely one map (natural transformation) $A \to 1$. Hence one can view a (generalised) \emph{point}, or \emph{element}, of $\Sigma$ as a map (natural transformation) $1 \to \Sigma$ and this map is precisely $\lambda$. Hence a global section of an object $A$ in $\mathcal{E}$ is a generalised element of $A$. In this language the Kochen-Specker theorem \cite{ks1}, or the existence of \emph{contextuality},  reads \cite{ib3}:
\begin{theorem} (Kochen-Specker, Isham-Butterfield)\label{t0}
For $\Sigma: \mathcal{V(H)} \to \mathsf{Sets}$, there exists no global section $1 \to \Sigma$,
\end{theorem} 
\noindent and hence $\Sigma$ has no points at all in a generalised sense! 
An alternative view presents itself. For sets, 
surjective maps are right cancellable; if $g_1 \circ f = g_2 \circ f$
then $g_1=g_2$. This provides the categorical version of a surjective map, called an \emph{epic} map, or an \emph{epimap} or an \emph{epimorphism}. Just as any function $A \to \{*\}$ is surjective, in any topos of presheaves the map $A \to 1$ is an epimap.
Hence one may view the section $1 \to \Sigma$ as the 
``cross-section"
(i.e., right inverse) of an epimap 
$\Sigma \to 1$; following standard terminology  we also call this map a section (indeed, this is the reason for calling $1 \to \Sigma$ a section in the first place). As shall become evident, the nonexistence of such a section is not only critical for a realist description of quantum theory, it is also of crucial importance to logic. Considering
any subposet $\mathcal{W(H)}$ of $\mathcal{V(H)}$ we may make the following definition:
\begin{definition}\label{def:sdc}
$\mathcal{W(H)}$ is \emph{contextual} if $\Sigma: \mathcal{W(H)} \to \mathsf{Sets}$ has no global section.
\end{definition}
This allows us to look for contextual ``scenarios", i.e., instances of contextuality, by considering a small number of 
observables, constructing the corresponding subposet of $\mathcal{V(H)}$, and looking for global sections of $\Sigma$ there
(we always refer to $\Sigma$ despite the different domains; the context should make clear the poset to which we refer).

Theorem \ref{t0} does not depend on any choice of (pure) state; there are numerous
examples of so-called ``state-dependent" proofs of the Kochen-Specker theorem which do invoke a particular state, physically interpreted as the impossibility of observables possessing values \emph{in that given state}. Such proofs can also be displayed in terms of (the lack of) global sections
of the ``pseudostate" $\mathfrak{w}^{\state{\varphi}}: \mathcal{V(H)} \to \mathsf{Sets}$ (or subposets of $\mathcal{V(H)}$ thereof) - this
provides a unified perspective of state-dependent and state-independent versions of the Kochen-Specker theorem.
With $P(V)$ denoting the projection lattice of $V$ (seen as a von Neumann algebra) and with $\varphi \in \mathcal{H}$ fixed,
we first define the presheaf $S_{\state{\varphi}}$ (neglecting the restrictions) by
\begin{equation}\label{eq:saus2}
S_{\state{\varphi}} (V):=\bigwedge \left \{ P \in P(V) ~|~ P \geq \state{\varphi}\dstate{\varphi} \right \},
\end{equation}
{\it viz.} the smallest projector $P \in P(V)$ which dominates $\state{\varphi}\dstate{\varphi}$.
The Gelfand transform $P \mapsto \bar{P}$, defined as $\lambda_V (P) = \bar{P}(\lambda_V)$, dictates that $\bar{P}$
can be viewed as the characteristic function on $\Sigma (V)$ of those $\lambda_V$ for which $\bar{P}(\lambda_V)=1$.
This motivates the definition of 
$\mathfrak{w}^{\state{\varphi}}$, given by 
\begin{equation}\label{eq:saus1}
\mathfrak{w}^{\state{\varphi}} (V) =\{\lambda_V \in \Sigma (V)\mid \lambda_V (P) = 1\},
\end{equation}
 which is manifestly a subset of 
$\Sigma (V)$ for each $V$, and hence $\mathfrak{w}^{\state{\varphi}}$ is a \emph{subobject} of $\Sigma$
(the restriction maps for $\mathfrak{w}^{\state{\varphi}}$ are of course the same as for $\Sigma$; $\mathfrak{w}^{\state{\varphi}} (U \subseteq V)(\lambda _V) = \lambda_U$).
Continuing to write $\mathcal{W(H)}$ for some fixed subposet of $\mathcal{V(H)}$, we have the following definition:
\begin{definition}\label{def:sdc}
State-dependent contextuality is present for $\left(\state{\varphi} \in \mathcal{H}, \mathcal{W(H)}\right)$ if there is no global section $1 \to \mathfrak{w}^{\state{\varphi}}$.
\end{definition}
The following simple observation demonstrates the relationship between state-dependent and state-independent contextuality:
\begin{proposition}\label{prop:gsr}
For any $\mathcal{W(H)}$, if $\Sigma$ has no global section then there does not exist a $\state{\varphi} \in \mathcal{H}$ for which $\mathfrak{w}^{\state{\varphi}}$ has a global section.
\end{proposition}
\noindent Since $\mathfrak{w}^{\state{\varphi}}$ picks out a subset of all possible sections above each $V \in \mathcal{W(H)}$, the proof is clear.  

We end this subsection by noting that connections between contextuality andglobal  sections have also been given elsewhere;
for instance sheaf theoretic methods in contextuality and non-locality have  been  employed to great effect by Abramsky and coworkers (e.g., \cite{abram1, abram2}), who demonstrate that contextuality can be displayed by an operational consideration of measurement outcomes, without recourse to the quantum formalism.

\subsection{Mermin's state-dependent proof}\label{subsec:merm}

We let $\mathcal{H}\equiv \mathcal{H}_1 \otimes \mathcal{H}_2 \otimes \mathcal{H}_3 := \mathbb{C}^2\otimes \mathbb{C}^2 \otimes \mathbb{C}^2$ with the Pauli operators 
$\sigma_x , \sigma_y , \sigma_z$ having the standard  definitions, and write $X_1 \equiv \sigma _x \otimes \id \otimes \id$, $Y_2 \equiv \id \otimes \sigma_y \otimes \id$, {\it etc}.
Consider the following putative valuation:
 
\begin{align}
 \lambda(X_1X_2X_3) =&+1 \label{ghz1} \\
 \lambda(X_1Y_2Y_3) =  &-1 \\
 \lambda (Y_1X_2Y_3) = &-1 \\
\lambda (Y_1Y_2X_3) = &-1. \label{ghz4}
\end{align}
By definition $\lambda$ is multiplicative on commuting operators (and strictly is defined on the algebras generated by the given qubit observables). The product of the four terms on the left hand sides of  \eqref{ghz1}-\eqref{ghz4}
is $1$, whereas the product of the four values on the right hand side is $-1$, thus ruling out the mapping $\lambda$ as a viable valuation. 

The values taken by $\lambda$ on the right hand side of \eqref{ghz1} - \eqref{ghz4} are precisely those given by (expectation values in)
the  GHZ state \cite{GHZ}:
\begin{equation}
\state{\Psi} \equiv 1 / \sqrt{2} \left(\state{000} + \state{111} \right)
\end{equation}
 (the $0$'s and $1$'s correspond to eigenvalues of $\sigma_z$; see \cite{mer2} for Mermin's state-dependent Kochen-Specker proof). The GHZ state is not multiplicative on the given observables and therefore does not satisfy the definition of a section. In order to prove
that we are in the presence of state-dependent contextuality in the language of this paper, i.e., in accordance with definition \ref{def:sdc}, it must be shown that
$\mathfrak{w}^{\state{\Psi}}$  has no global section, i.e., that the local sections, compatible with the constraints
imposed by the GHZ state, cannot be appropriately ``glued" to give a global section with the required restrictions.

Let $\tilde{\lambda}: 1 \to \mathfrak{w}^{\state{\Psi}}$  denote a hypothetical global section of 
$\mathfrak{w}^{\state{\Psi}}$, and define $V_1$ to be the observable algebra generated by $\{ X_1, X_2, X_3 \}$ and, following equations \eqref{ghz1} - \eqref{ghz4}, {\it mutatis mutandis}, do the same for $V_2, V_3$ and $V_4$. We denote by $V_{X_1}$ the algebra generated by $X_1$ (which is contained in $V_1$), {\it etc}. We have that $\tilde{\lambda}_{V_i}:\{* \} \to \mathfrak{w}^{\state{\Psi}}(V_i)$ so that $\tilde{\lambda}_{V_i}(\{ * \})$ is a local 
section of $\mathfrak{w}^{\state{\Psi}}$ above $V_i$ for $i \in \{1...4 \}$. Hence we have the constraints:
\begin{align}\label{eq:secp}
\tilde{\lambda}_{V_1}(*)(X_1)\tilde{\lambda}_{V_1}(*)(X_2)\tilde{\lambda}_{V_1}(*)(X_3)=&+1 \nonumber\\
\tilde{\lambda}_{V_2}(*)(X_1)\tilde{\lambda}_{V_2}(*)(Y_2)\tilde{\lambda}_{V_2}(*)(Y_3)=& -1 \\
\tilde{\lambda}_{V_3}(*)(Y_1)\tilde{\lambda}_{V_3}(*)(X_2)\tilde{\lambda}_{V_3}(*)(Y_3) = & -1 \nonumber \\
\tilde{\lambda}_{V_4}(*)(Y_1)\tilde{\lambda}_{V_4}(*)(Y_2)\tilde{\lambda}_{V_4}(*)(X_3) = & -1. \nonumber
\end{align}
For convenience we write $\lambda_{V_i} := \tilde{\lambda}_{V_i}(*)$. The local sections $\lambda_{V_i}$ 
have restrictions, for example $\lambda_{V_1}|_{V_{X_1}} = \lambda_{V_{X_1}}$ etc. It is thus required
that the compatibility conditions hold: $\lambda_{V_1}|_{V_{X_1}}=\lambda_{V_2}|_{V_{X_1}}$, $\lambda_{V_1}|_{V_{X_2}}=\lambda_{V_3}|_{V_{X_2}}$
and so on. These are manifestly requirements of non-contextuality. Appropriate substitution back into \eqref{eq:secp}
gives rise to the same contradiction found on $\lambda$ in \eqref{ghz1} - \eqref{ghz4}. The following Hasse diagram illustrates the poset structure of the GHZ scenario:
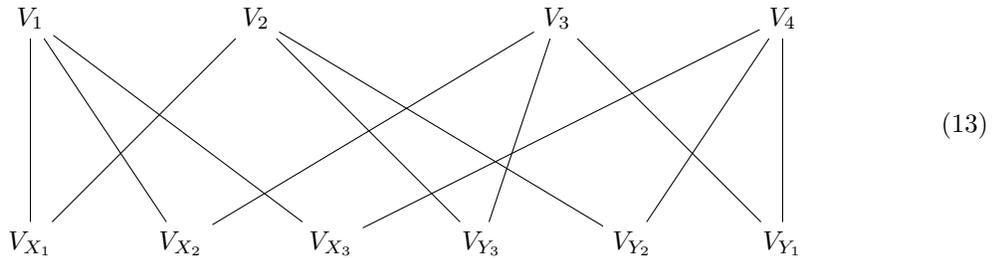
\begin{small}
\begin{center}
\begin{equation}\label{d:ghz} 
\begin{tikzpicture}[baseline=(current  bounding  box.center)]

  \node (g1) at (-5,3) {$V_1$};
  \node (g2) at (-2,3) {$V_2$};
  \node (g3) at (2,3) {$V_3$};
   \node (g4) at (5,3) {$V_4$};
 
  \node (x1)at (-5,0) {$V_{X_1}$};
  \node (x2) at (-3,0) {$V_{X_2}$};
  \node (x3) at (-1,0) {$V_{X_3}$};
 \node (y3) at (1,0) {$V_{Y_3}$};
  \node (y2)at (3,0) {$V_{Y_2}$};
  \node (y1)at (5,0) {$V_{Y_1}$}; 
 % \node (min) at (0,-2) {$\{\}$};

 \draw  (g1)--(x1);
  \draw  (g1)--(x2) ;
  \draw  (g1)--(x3) ;

 \draw  (g2)--(x1) ;
  \draw  (g2)--(y2);
  \draw  (g2)--(y3);

 \draw  (g3)--(x2);
  \draw  (g3)--(y3) ;
  \draw  (g3)--(y1);

  \draw  (g4)--(y2) ;
  \draw  (g4)--(x3);
  \draw  (g4)--(y1);

 % \draw[preaction={draw=white, -,line width=6pt}] (g1) -- (x2) -- (g3);
\end{tikzpicture}
\end{equation}
\end{center}
\end{small}

\section{Measurement Based Quantum Computation}\label{sec:mbqc}
In this section $\ell$-2 MBQC is introduced and its connection with contextuality is emphasised. The main result is theorem
\ref{MBQC_C}, which demonstrates the necessity of contextuality for non-linear function evaluation, the simplest manifestation
of which is the contextual OR-gate of Anders and Browne \cite{ab1} constructed from Mermin's state-dependent Kochen Specker proof.

\subsection{$\ell$-2 MBQC}
Measurement based quantum computation (MBQC) employs single-particle (qubit or qudit, projective or positive operator valued) measurements on a fixed initial state to achieve a computational output. The state is altered by the local measurements (in particular any entanglement is destroyed); the initial state is thus viewed as a resource, and for suitable choices such as cluster states \cite{Cluster1, Cluster2}, MBQC is universal for quantum computation. 

The measurements generating the computation have individually random but correlated outcomes, and the computational result depends crucially on these correlations.  In addition, the observables to be measured must typically be adjusted according to previous measurement outcomes. Therefore, classical side-processing is required both to direct the computation and to produce the desired output. This classical processing is typically limited, and is not universal for classical computation. For example, in the universal scheme of \cite{RB01}, only addition modulo 2 is permitted. 

We note that universal MBQC schemes with different classical side-processing, different resource states and different sets of allowed measurements have been devised; see, e.g., \cite{Leung, GE, CDJZ, Miy, WAR, ElseBartlett}. However, for the present purpose, we consider MBQC on quantum states of multiple qubits, where the classical side-processing is restricted to addition modulo 2, see \cite{RB01, ab1, rau1}. 

The relevant Hilbert space for our work is $\mathbb{C}^{2^n} \cong \bigotimes_{k=1}^n \mathbb{C}^2$ and measured observables are non-trivial on only one factor in the tensor product; these are referred to as \emph{local observables}. Local measurements are defined in the obvious way. We now give a brief definition of $\ell$-2 MBQC and refer to
\cite{rau1} for details; for clarity of exposition we restrict attention to the case where there is only one output bit, though
our results pertain to the general case. We also restrict ourselves to computations which are \emph{temporally flat}, i.e., 
joint measurements of local observables or, in other words,
where the choice of measured local observable does not depend on the outcome of a previous measurement.

\begin{definition}\label{def:l2}
A temporally flat, deterministic $\ell$-2 MBQC $\mathcal{M}$ consists of a resource state $\state{\Phi} \in \mathbb{C}^{2^n}$, classical input $i \in Z_2 ^m$ and classical output $o : Z_2 ^m \to Z_2$, $i \mapsto o(i)$, a collection
of (maximal) local observables $\{O_k({q_k}) |k \in \{1...n \}; q_k \in Z_2 \}$ 
for which the measurement outcomes of a given $O_k({q_k})$ are labelled $s_k({q_k}) \in \{0,1\}$ for each $k$.
The computed output $o(i)=\sum_{k} s_k({q_k})$ mod 2, and, with $q=(q_1...q_n),$ the measured observable $O_k({q_k}) $ is related to the outcome
$ s_k({q_k}) $ by $q=Qi$ where $Q:Z_2 ^m \to Z_2 ^n$.
\end{definition}
Since all the computations we consider are of the above form, for brevity we occasionally omit the qualifiers ``deterministic" and ``temporally flat", and only refer to such conditions when necessary. 

Given $B(\mathbb{C}^{2^n})$, we must construct the poset $\mathcal{W}(\mathbb{C}^{2^n})$, yielding
a diagram similar to \eqref{d:ghz}. 
We proceed by considering all strings $\{(O_1(q_1),..., O_n(q_n))\}$
for which $O_1 (q_1) ... O_n(q_n) \state {\Psi} = (-1)^{o(i)}\state{\Psi}$. Given such a string $\left( O_1(q_1),..., O_n (q_n) \right)$ the Abelian algebra (context) is generated by the local observables, written 
$\{O_1(q_1),..., O_n(q_n) \}^{\prime \prime}$ (the prime denotes the commutant; see, e.g., \cite{kring1} theorem 5.3.1.). 

\begin{definition}
Let $\{ O_i\}$ denote the collection of strings of local observables given above with $O_i \cap O_j$ the intersection of any pair,
giving morphisms $O_i \cap O_j \to O_i$ and $O_i \cap O_j \to O_j$, etc, and call this poset $\mathcal{W(H)}_0$. Then $\mathcal{W(H)}$ is the poset defined by taking the algebra generated by each object in $\mathcal{W(H)}_0$.
\end{definition}

Hence the collection $\{O_i\}$ forms the ``top layer" of the poset, with the second layer given as the algebras generated by 
elements in the intersection of the generating sets of the maximal algebras, {\it etc.} (note that any  $\{O_1(q_1),..., O_n(q_n) \}^{\prime \prime}$ is maximal since $O_1 (q_1) ... O_n(q_n)$  has $2^n$ mutually orthogonal rank-1 projections in its spectral resolution,
and the algebra generated by these projections coincides with $\{O_1(q_1),..., O_n(q_n) \}^{\prime \prime})$. A maximal Abelian subalgebra is called a \emph{masa}. The main result of \cite{rau1} reads:
\begin{theorem}\emph{ \cite{rau1}}\label{MBQC_C} Every $\ell$-2 MBQC which deterministically computes a non-linear Boolean function is contextual.
\end{theorem}
The version of contextuality used in \cite{rau1} is the ``strong contextuality" of \cite{abram1}, which, though different from our definition, coincides with it in the cases we consider.

We conclude this subsection with two remarks. Firstly, there are quantum algorithms with super-polynomial speed-up in the class of deterministic MBQCs, exhibiting contextuality. Specifically, the quantum algorithm for the ``discrete log'' problem \cite{Shor} breaks a crypto system. It is deterministic in the circuit model \cite{ZaMos}, and can be converted into deterministic MBQC by standard techniques \cite{RB01}. Secondly, the condition of determinism in theorem \ref{MBQC_C} can be relaxed. For probabilistic function evaluation, for every given non-linear function there is a threshold in the success probability above which the corresponding MBQC is contextual. This threshold can be low; for the bent functions \cite{MWS} it approaches 1/2 in the limit of large number of input bits \cite{rau1}.

\subsection{Anders and Browne's OR-gate}\label{subsec:AB}

The first and simplest example that illustrates the connection between contextuality and non-linear function evaluation in MBQC is the OR-gate of Anders and Browne \cite{ab1}, which is directly constructed from Mermin's state-dependent proof of the Kochen-Specker theorem.

In order to consider computation, it is necessary to rewrite the proof of contextuality arising from the GHZ state 
in computational variables. Consider once more the ``attempted GHZ-valuation" $\lambda$ given in \eqref{ghz1} - \eqref{ghz4}.
Define $\lambda (X_i)=(-1)^{x_i}$ and $\lambda(Y_i)=(-1)^{y_i}$. Direct substitution into \eqref{ghz1} - \eqref{ghz4} yields the following constraints (with all sums modulo 2):
\begin{equation}\label{MermEq}
\begin{array}{rcl}
x_1+x_2+x_3  &=&0,\\
x_1+y_2+y_3  &=&1,\\
y_1+x_2+y_3  &=&1,\\
y_1+y_2+x_3  &=&1,
\end{array}
\end{equation} 
resulting, once again, in an insoluble set of equations (this time seen by summing the left hand side of \eqref{MermEq}, summing the right hand side, and comparing).
The above proof of the Kochen-Specker theorem can be adapted to MBQC. There is no contradiction if in
\eqref{ghz1} - \eqref{ghz4}, or equivalently their computational counterparts in \eqref{MermEq}, the $x_i,y_i$ are regarded as actual measurement outcomes rather than predetermined (noncontextual) values.

The right hand side of (\ref{MermEq}) can be viewed as the output of an OR-gate in $\ell$-2 MBQC (recall definition \ref{def:l2}) as follows. The input
$i = (i_1, i_2) \in Z_2 \times Z_2$ and $o(i) \in Z_2$. If $q_k=0$, $X_k$ is measured; if $q_k=1$, $Y_k$ is measured. The outcomes
$s_k({q_k})$ are written $s_k{(0)}=x_k$, $s_k(1)=y_k$.
 The map $Q$ effects $q_1=i_1$, $q_2=i_2$, $q_3=i_1+i_2$ mod 2. Hence $o(i)=s_1({q_1})+s_2({q_2})+s_3({q_3})$ so,
for instance, if $(i_1, i_2)=(0,0)$, $q(0,0) = s_1{(0)}+s_2(0)+s_3(0) = x_1+x_2+x_3 = 0$. Computing the outputs for the remaining inputs verifies that $o(i_1, i_2)=OR(i_1, i_2)$, and, given equation \eqref{MermEq}, we see how the above Kochen-Specker proof gives rise to an OR-gate in $\ell$-2 MBQC.

The OR-gate has the property of being a non-linear Boolean function (compare $OR(1,1) + OR(0,1)$ with $OR(1,0)$) which, as theorem \ref{MBQC_C} shows,
implies the presence of contextuality.
While not of practical use, the ability to evaluate an OR-gate within MBQC is of fundamental relevance. The classical control computer in $\ell$-2 MBQC is limited to addition mod 2, and is thus by itself not classically universal. Access to a GHZ state  and Pauli measurements  promotes it to classical universality, vastly increasing its computational power.

\section{Topoi and logic}\label{sec:tal}

In this section some basic elements of topos theory are given, with particular emphasis on logic. The subobject classifier is defined and the result that the topos version of the axiom of choice implies Booleanness is stated.

Topos theory emerged from two distinct places (a good history can be found in the introduction of \cite{jon1}): Grothendieck's work on sheaves and Lawvere and Tierney's work on categorical set theory and the foundations of mathematics; it is here 
that the logical aspects of topos theory are most visible. 

A topos resembles a ``universe of sets"; certain universal constructions available for sets such as Cartesian products, terminal objects, exponentials ($B^A :=\{f\mid f:A \to B \} $ is a set) and so on, are available in any topos. However, most topoi 
are non-Boolean, manifested by the failure of the excluded middle, and therefore constructive arguments must be used. Moreover, AC fails in general and, as we shall see, this also has an impact upon the logic of the topos.

\subsection{Logic and Heyting algebras}

The motivating example of a Heyting algebra is the lattice $\mathcal{O}(X)$ of open sets of a topological space $X$. 
Interpreting the open sets as propositions, the logical ``and" ($\wedge$) and ``or" ($\vee$) are, just as for a Boolean algebra, set theoretic intersection ($\cap$) and union ($\cup$), respectively, since these operations preserve openness. However, logical negation $\neg$ cannot be interpreted as set-theoretic complementation as the complement fails to be open in general, and rather $\neg U :=int U^c$ ($c$ to denote set-theoretic complement.) Any derived operation involving
$\neg$ must also be modified. The lattice of open sets of a topological space, under these operations, forms a \emph{Heyting algebra}. However, there are Heyting algebras not of this form, and so we provide the general definition.

\begin{definition}
A Heyting algebra $\mathfrak{H}$ is a bounded distributive lattice (like a Boolean algebra) with a top element 1, a bottom element 0, and a binary operation
$\implies$ which satisfies, for all $a,b,c$ in $\mathfrak{H}$, $c \leq a \implies b$ if and only if $(c \wedge a) \leq b$.
\end{definition} 
The negation $\neg a$ is defined by $a \implies 0$ and is called the \emph{pseudo-complement} (of $a$); it is the greatest element $b$
of $\mathfrak{H}$ for which $a \wedge b= 0$. An element
$a \in \mathfrak{H}$ is \emph{regular} if $\neg \neg a = a$ and is \emph{complemented} if the pseudo-complement satisfies
\begin{equation}
a \vee \neg a =1;
\end{equation}
in general $a \vee \neg a \leq 1$ and thus the excluded middle is not available in a general Heyting algebra. This is the principal difference between Heyting algebras and Boolean algebras. Indeed (see \cite{bor3}, proposition 1.2.11 and the accompanying proof),
\begin{proposition}\label{prop:hb}
A Heyting algebra $\mathfrak{H}$ is Boolean if, and only if, for each $a \in \mathfrak{H}$, $\neg \neg a = a$, i.e., if each element is regular which holds if and only if for all $a \in \mathfrak{H}$, $a \vee \neg a = 1$ (i.e., each element is complemented).
\end{proposition} 

Heyting algebras arise in topos theory in governing both the structure of subobjects of a given object, and the structure of the global elements 
of the subobject classifer.

\subsection{The topos $\mathsf{Sets}$}

The category $\mathsf{Sets}$ of sets and functions provides an archetypal topos.
The logical structure of $\mathsf{Sets}$ encompasses how subsets (of a given set) relate to each other. Given $S \subseteq A$, the categorical
``philosophy" dictates that one should rephrase this only in terms of functions. One can view the subset relation $S \subseteq A$ alternatively as an injective map $S \rightarrowtail A$ (or, more properly, an equivalence class of injective maps). These maps are ``classified" by characteristic functions $A \to \{0,1\}$, i.e., given any $s:S \rightarrowtail A$ there is a unique
map $\chi_S$ whose inverse image $\chi_S^{-1}(1) = s(S)$. Thus with $subA$ denoting the collection of injective maps into $A$, $\mathcal{P}(A)$ the power set  of $A$ and  $2^A$ the collection of all functions $A \to 2$,
\begin{equation}\label{eq:sub2}
subA \cong \mathcal{P}(A) \cong 2^A.
\end{equation}

As a case in point, consider $\Sigma : V \to \mathsf{Sets}$, where $V$ is a finite Abelian algebra. Then $\Sigma (V)$ is the finite analogue of a classical phase space; $sub(\Sigma)$ is a Boolean algebra corresponding to subsets of the phase space,
and a pure state is either in or out of a given subset, as witnessed by the ``truth value set" $2 =\{0,1\}$.

It is possible to give an analogous correspondence between subobjects of a given object $A$ and maps $A \to \Omega$
in any topos, where $\Omega$ represents a ``truth value object" generalising $2$.
The required definitions can be given without reference to points, which is crucial given that objects in a topos  may have many subobjects but no points at all (e.g., $\Sigma$ in the presence of contextuality).

\subsection{General topoi}

The full definition of a topos, which is technical, will not be required
(see, e.g., \cite{bor3} ch. 5). Instead we provide only the concepts necessary for the definition of the subobject classifier which embodies the logical structure of a topos, and discuss the relationship between the logic and the topos formulation of AC. The appropriate generalisation of subset to subobject is given
as a \emph{monic} (left-cancellable) map: 
\begin{definition}
A subobject $A$ of an object $B$ in $\mathcal{E}$ is defined as a monic arrow $A \rightarrowtail B$.
\end{definition}
Of course, ``monic" is the categorical version of  ``injective" in $\mathsf{Sets}$.
The definition of subobject classifier in a topos $\mathcal{E}$ mirrors the correspondence in $\mathsf{Sets}$ between
subsets and functions with co-domain $2$:
\begin{definition}\label{def:subobj}

Let $\mathcal E$ be a topos. A {\emph subobject classifier} is an object $\Omega$ together with a morphism $\mathrm{true} :1\rightarrow \Omega$
such that for every object $A$ of $\mathcal E$ and every monic $s: S \rightarrowtail A$, there is a unique morphism $\chi_S: A\rightarrow \Omega$ such that
the following square is a \emph{pullback} (i.e., commutes and is universal among
such squares):

\begin{equation}\label{eq:subc}
\begin{diagram}
\node{S}\arrow{e,t}{}\arrow{s,l,V}{s}\node {1}\arrow{s,r}{true}\\
\node{A}\arrow{e,b}{\chi_S}\node{\Omega}
\end{diagram}\hspace{40pt}
\end{equation}
\end{definition}

This sets up a Heyting algebra isomorphism between subobjects of some object $X$ and characteristic morphisms $X \to \Omega$;
\begin{equation}
subX \cong Hom(X, \Omega) \cong \Omega^{X}
\end{equation}
where $Hom(X,\Omega)$ denotes the set of maps (morphisms) $X \to \Omega$. We state without proof the following fact, referring the reader to \cite{bor3}, proposition 6.2.1:
\begin{proposition}
For any object $X$ in $\mathcal{E}$, $subX$ is a Heyting algebra.
\end{proposition}
For example, the collection of subobjects of $\Sigma: \mathcal{V(H)} \to \mathsf{Sets}$ forms a Heyting algebra which is not Boolean,
giving the topos version of phase space a fundamentally different structure from that which is usual for classical systems.

 Replacing $X$ by $1$---the terminal object in $\mathcal{E}$---immediately yields
\begin{equation}\label{eq:omega1}
sub(1) \cong \Gamma \Omega ,
\end{equation}
demonstrating at once that the global elements of $\Omega$ have the structure of a Heyting algebra, given as the subobjects of the terminal object $1$.

\begin{definition}
$\mathcal{E}$ is called \emph{Boolean} if $\text{\emph{sub}}\Omega$ (the Heyting algebra of subobjects of $\Omega$) is a Boolean algebra.
\end{definition}

\begin{definition}\label{d1}
 If every epimap in $\mathcal{E}$ has a section, then we say the AC holds in $\mathcal{E}$.
\end{definition}
This is a straightforward generalisation from the standard scenario (of set theory) where it is usually maintained that 
AC holds. AC fails in ``most" topoi, and has a surprising connection to logic. 

\begin{theorem} \label{th:dia1}\emph{(Diaconescu)}.
In a topos $\mathcal{E}$ if AC holds then $\mathcal{E}$ is Boolean.
\end{theorem}
Though the converse fails in general, in any (\emph{localic}) topos, for example the topos of presheaves on a poset, the converse does hold (see \cite{bor3}, proposition 7.5.5 or theorem \ref{t1} below). Thus we may tie the non-existence of
a global section of $\Sigma \to 1$ or $\mathfrak{w}^{\state{\varphi}} \to 1$, and hence contextuality, to non-Boolean topos logic.

\section{Presheaves on a poset and $\widehat{\mathcal{W(H)}}$}\label{sec:prspos}

In this section we present some basic facts about topoi of presheaves on a poset, and provide a construction for quantifying
the non-Booleanness of the subobject classifier and therefore the topos as a whole. The quantum topos $\widehat{\mathcal{W(H)}}$ is introduced and physically motivated as representing a subtopos of the Isham-Butterfield topos
$\widehat{\mathcal{V(H)}}$.

\subsection{General case}
Let $(P, \leq)$ denote a poset, henceforth written $P$.
The following demonstrates
that $\widehat{P}$ is Boolean if and only if $P$ consists of only one point:
\begin{proposition}\label{prop:grpd}  \emph{({\cite{ele1}}, lemma 1.4.12.)}
Let $\mathcal{C}$ be an arbitrary (locally small) category. $\widehat{\mathcal{C}}$ is Boolean if and only if $\mathcal{C}$ is a groupoid.
\end{proposition} 
\noindent Thus it is trivial to observe when a given topos of presheaves on a poset is Boolean. If $\mathcal{C}$ were just a single Abelian algebra (a poset with one point), $\widehat{\mathcal{C}} \cong \mathsf{Sets}$, i.e., one recovers the topos of sets and functions.
However, we also wish to quantify the non-Booleanness of a presheaf topos on a poset, and to do this we first view $\widehat{P}$ as a topos of \emph{sheaves} on a topological space. This equivalence also allows for the demonstration that contextuality gives rise to non-Boolean logic.

\begin{definition}
 A \emph{down-set} $A$ on a poset $P$ is a subset $A \subseteq P$ for which $x \in A$ and $y \leq x$ implies $y \in A$.
\end{definition}
The down-sets $\downarrow P$ on $P$ define the open sets of the \emph{anti-Alexandrov topology} on $P$; we write $\mathsf{P}$ for this space. 

\begin{definition}
Let $X$ be a topological space and $\mathcal{F}: \mathcal{O}(X) \to \mathsf{Sets}$ a presheaf on the open sets $\mathcal{O}(X)$. $\mathcal{F}$ is a \emph{sheaf} if for any open set $U$ and any open cover $U=\cup_iU_i$ of $U$ and each collection
$f_i \in \mathcal{F}(U_i)$ (of local sections) for which $f_i|_{U_i \cap U_j}=f_j|_{U_i \cap U_j}$, there exists a unique 
$f \in \mathcal{F}(U)$ such that $f|_{U_i}=f_i$ for each $i$.
\end{definition}
\noindent We denote the collection of (set-valued) sheaves on $X$ by $Sh(X)$.
It turns out (e.g., \cite{ele2}) that the presheaves on $P$ seen as a poset are exactly the sheaves on $\mathsf{P}$:
\begin{proposition}\label{prop:shc}
 $\widehat{P} \cong Sh(\mathsf{P})$ where $\cong$ denotes isomorphism of categories.
\end{proposition}

Furthermore (see {\cite{mor1}, ch. II, proposition 4), with $1$ denoting the terminal object in $Sh(\mathsf{P})$ and with  \eqref{eq:omega1}  and proposition \ref{prop:shc},
\begin{equation}\label{eq:shboo}
\mathcal{O}(\mathsf{P}) \cong \downarrow P \cong sub (1) \cong \Gamma \Omega
\end{equation}

This enables us to understand the logical structure of the truth values $\Gamma \Omega$ in $\widehat{P}$ by directly looking at the poset $P$, which will be of use when we consider the topoi involved in quantum computation. Since $sub B$ is a Heyting algebra for any $B$ in $\mathcal{E}$, the complement of a subobject $A \rightarrowtail B$
is defined as the Heyting algebra complement, and the existence of complements is strongly tied to the logic.

\begin{theorem}\label{t1}
Let $\mathcal{E}$ be a topos of the form $\mathcal{E} = Sh(X)$ for $X$ a topological space. The following are equivalent \emph{(\cite{bor3}, proposition 7.5.5, and \cite{jon1}, proposition 5.14)}:
\begin{enumerate}
\item $\Omega$ is a Boolean Algebra (i.e., $\mathcal{E}$ is Boolean);
\item AC holds in $\mathcal{E}$;
\item All subobjects have complements ;
\item $\mathcal{O}(X)$ is a complete Boolean algebra. \label{item4}
\end{enumerate}
\end{theorem}

One can see, for example, that when defined on the poset \ref{d:ghz}, $\mathfrak{w}^{\state{\varphi}} \rightarrowtail \Sigma$ has no complement, showing the non-classical structure of the phase space $\Sigma$. Theorem \ref{t1} also highlights the interrelatedness of many of category/topos theoretic concepts: non-Booleanness arises from the (failure of) AC, or via the lack of complements of subobjects, or by the failure of the base category to be a groupoid.

However, we wish to not only
indicate whether a topos is non-Boolean, but to be able to say \emph{how} non-Boolean it is. Equation \eqref{eq:shboo}
shows that $\Gamma \Omega$ is Boolean if and only if $\mathcal{O}(\mathsf{P})$ is, and  by theorem \ref{t1}, this is equivalent to $\widehat{P}$ being Boolean. Hence the (non-)Booleanness of $\Gamma \Omega$ governs the (non)-Booleanness of $\widehat{P}$.

Since 
we aim to show that non-Boolean logic is a resource for quantum computation, the quantification of such a resource is crucial, and making such a notion precise is the subject of the next subsection.

\subsection{Quantifying non-Booleanness}
In light of the fact that $\widehat{P}$ is Boolean precisely when the Heyting algebra $\Gamma \Omega$ is, which by proposition \ref{prop:hb} is the case precisely when each element of $\Gamma \Omega$ is complemented, 
the non-Booleanness of $\Gamma \Omega$ can be captured as the ratio of non-complemented elements
in $\Gamma \Omega$ to the total number of elements. In turn, this serves to indicate the non-Booleanness of $\widehat{P}$.

We write $\text{comp}(\Gamma \Omega)$ for the sub-Heyting algebra of $\Gamma \Omega$ consisting of complemented elements, which is also a Boolean algebra, and write $|\text{comp}(\Gamma \Omega)|$ for its cardinality.

\begin{definition}
The \emph{non-Booleanness} $\mathfrak{q}$ is given by
\begin{equation}
\mathfrak{q} \Gamma \Omega = 1- \frac{|\text{\emph{comp}} \Gamma \Omega|}{ |\Gamma \Omega|}
\end{equation}
\end{definition}

Thus, if all global elements of $\Omega$ are complemented, $\mathfrak{q} \Gamma \Omega = 0$. For a poset of fixed size,
$\mathfrak{q}$ therefore defines a monotone function with range $\left[0,1-\frac{2}{|\Gamma \Omega|}\right]$. Only in the limit $|\Gamma \Omega| \to \infty$ can $\mathfrak{q}\Gamma \Omega$ become $1$.

\subsection{The quantum topoi}

Only in the last 20 years or so has topos theory been applied to problems in the foundations of quantum mechanics; topos quantum theory has attracted a large number of contributions in recent years, most focussing on the role of the many classical ``pictures" embedded in the quantum description (e.g. \cite{heuclas} and references therein).

The topos $\mathsf{Sets}$ captures the logical structure of classical physics, where Boolean logic plays a fundamental role both in the structure of propositions (i.e., distinguished subsets of the phase space) and in the truth evaluation
of the given propositions - any such proposition is true or false, and the two truth values are arranged as a Boolean algebra.
It has been argued by Isham and collaborators that one should use topoi other than $\mathsf{Sets}$
for construction of physical theories in general. Isham and D\"{o}ring in \cite{id1} provide a thorough discussion of the topos $\widehat{\mathcal{V(H)}}$ (which we recall is the topos of presheaves on the poset of Abelian von Neumann subalgebras of $B(\mathcal{H})$). The spectral presheaf $\Sigma$ carries a logical structure of its own and represents the phase space 
in the topos $\widehat{\mathcal{V(H)}}$.

The topos $\widehat{\mathcal{W(H)}}$ of interest to us is a subtopos of  $\widehat{\mathcal{V(H)}}$ and these topoi have many features in common. We may view $\widehat{\mathcal{W(H)}}$ as the restriction to a specific physical situation of interest, with the given logical structure as the logic governing that situation.

Consider once more the poset $\mathcal{W(H)} \subseteq \mathcal{V(H)}$ and the topos
 $\widehat{\mathcal{W(H)}}$.
Combining definition \ref{d1} with theorem \ref{t1} we thus immediately see that non-Booleanness arises from sectionless epimaps; we thus have as a corollary of theorem \ref{t1} the following observation:
\begin{corollary}
Let $\mathcal{E}=\widehat{\mathcal{W(H)}}$ and let $\Sigma$ and $\mathfrak{w}$ have the given definitions. If $\Sigma$ has no global section or if there exists a  $\state{\varphi} \in \mathcal{H}$ for which $\mathfrak{w}^{\state{\varphi}}$ has no global section, then the logic of $\mathcal{E}$ is non-Boolean.
\end{corollary}\label{cor:cinb}
Therefore,
\begin{corollary}\label{c2}
State-dependent or state-independent contextuality implies non-Boolean logic in $\mathcal{E}$.
\end{corollary}

Thus state-dependent or independent contextuality may be viewed as a sufficient condition for the 
given topos to be non-Boolean, thus demonstrating 
a  link between the Kochen Specker theorem and the logic of a topos closely related to a concrete \emph{physical situation}. We now explore the connection between the non-classical logic of a topos and the promotion of non-universal to universal computation in MBQC.

\section{Non-Boolean logic and quantum computation}\label{sec:logcomp}

Before analysing the role played by non-Boolean topos logic in the class of $\ell$-2 MBQC of interest, we briefly summarise
the various topics that have been presented thus far, and how they are related. The state-dependent and state-independent Kochen-Specker theorems were rephrased
in terms of the non-existence of global sections of the presheaves $\mathfrak{w}^{\state{\varphi}}$ and $\Sigma$ respectively, both defined on a poset $\mathcal{W(H)} \subseteq \mathcal{V(H)}$, the latter representing the collection
of Abelian von Neumann subalgebras of $B(\mathcal{H})$, ordered by inclusion, and we presented Mermin's state-dependent proof
in this language. Temporally flat deterministic $\ell$-2 MBQC was introduced, with emphasis on a result in \cite{rau1} that
demonstrates the necessity of contextuality in the evaluation of non-linear Boolean functions, and we presented the simple example
of Anders and Browne's OR-gate \cite{ab1}. 

The connection to logic was then made by introducing some basic machinery of topos theory, where the Boolean logic
of set theory is replaced by (the non-Boolean) ``intuitionistic logic", with Heyting algebras governing the structure of the multivalued truth object and the structure of subobjects of the ``phase space" $\Sigma$.
The special case of (the topos of) presheaves on a poset, which encompasses the quantum case, was then analysed in greater detail, and we delineated conditions under which such a topos is Boolean. Via the topos version AC, we then established the major link between contextuality and logic: by constructing
the topos $\widehat{\mathcal{W(H)}}$ from the poset $\mathcal{W(H)}$, we showed (corollary \ref{c2}) that contextuality implies non-Boolean
logic in $\widehat{\mathcal{W(H)}}$. 

We are therefore presented with a link between non-Boolean logic and computation. With $\Phi$ a resource state for
an $\ell$-2 MBQC $\mathcal{M}$  as defined in section \ref{sec:mbqc}, the main result of  \cite{rau1}, in the language of this paper, reads
\begin{proposition}
If $\mathcal{M}$ computes a non-linear Boolean function, then $\mathfrak{m}^{\state{\Phi}}$ has no global section, i.e., 
there is state-dependent contextuality for $\left( \state{\Phi}, \mathcal{W(H)} \right)$. 
\end{proposition}

Hence we immediately arrive at one of our main results:
\begin{theorem}
If $\mathcal{M}$ computes a non-linear Boolean function then $\widehat{\mathcal{W(H)}}$ is a non-Boolean topos.
\end{theorem}
In other words, \emph{non-classical logic is a necessary condition for non-linear function evalutation, and hence universal classical computation} in $\ell$-2 MBQC, the model of computation
we consider. It does not seem unwarranted to view the non-Boolean logic of $\widehat{\mathcal{W(H)}}$ as affording a new potential ``resource" for quantum computation. Given other qualities previously mooted as computational resources---entanglement of $\state{\Phi}$ for instance---we may view non-Booleanness in the same light - a view which is to be strengthened in subsection \ref{subsec:cons} where we will demonstrate the ``consumption" of 
the non-Booleanness by the execution of the Anders and Browne OR-gate. Just as the entanglement is ``used up" in each step of the computation, yielding a product state after the computational output has been produced, we will show that the non-Booleanness follows a similar trajectory, with the ``final topos" being a Boolean one.

We turn attention once more to the example of Anders and Browne's OR-gate, in order to demonstrate explicitly the consumption of non-Booleanness in the course of the computation, and how this is correlated to the decrease in computational power at each stage. Our main conclusions, however, pertain also to any MBQC satisfying definition \ref{def:l2}.

\subsection{Non-Booleanness as a computational resource}\label{subsec:cons}

We again refer to the Anders-Browne OR-gate, and consider the situation after each measurement with regards to the non-Booleanness and computational power. The initial state is $\state{\Psi} = \frac{1}{\sqrt{2}}\left( \state{000} + \state{111}\right)$, with $\mathcal{W(H)}$ diagram \ref{d:ghz} and associated topos $\widehat{\mathcal{W(H)}}$.

Due to the fact that $(\state{\Psi}, \mathcal{W(H)})$ computes an OR-gate, all Boolean functions can be evaluated.
The quantity 
$|\Gamma \Omega|$ for the topos $\widehat{\mathcal{W(H)}}$ can be found by counting the down-sets in diagram \ref{d:ghz}, giving $|\Gamma \Omega| = 113$. The only
complemented elements are $0$ and $1$ (i.e., the empty set and the whole space), yielding $\mathfrak{q}\Gamma \Omega = 111/113$.

Without loss of generality we assume that $X_1$ is measured and the $+1$ outcome is registered, resulting in the state transformation $\state{\Psi} \mapsto \state{\Psi ^{\prime}}$, with
\begin{equation}
\state{\Psi ^{\prime}} = \state{+}\otimes (\frac{\state{00} + \state{11}}{\sqrt2}).
\end{equation}
This is the new resource state for the rest of the computation. It also specifies a subposet 
$\mathcal{W(H)}^{\prime} \subset \mathcal{W(H)}$: for deterministic computation in line with definition \ref{def:l2} the local observables are specified by
\begin{align}\label{eq:1m}
&X_1X_2X_3 \state{\Psi ^{\prime}} = \state{\Psi ^{\prime}}\\
\nonumber &X_1Y_2Y_3 \state{\Psi ^{\prime}} = -\state{\Psi^{\prime}},
\end{align}
i.e.,  we are restricted to the subposet defined by $V_1$ and $V_2$:

\begin{small}
\begin{center}
\begin{equation}\label{d:ghz3} 
\begin{tikzpicture}[baseline=(current  bounding  box.center)]

  \node (g1) at (-5,3) {$V_1$};
  \node (g2) at (-2,3) {$V_2$};
  
  \node (x1)at (-5,0) {$V_{X_1}$};

 \draw  (g1)--(x1);

 \draw  (g2)--(x1) ;

 % \draw[preaction={draw=white, -,line width=6pt}] (g1) -- (x2) -- (g3);
\end{tikzpicture}
\end{equation}
\end{center}
\end{small}

The existence of a valuation $\nu$ compatible with the constraints \eqref{eq:1m} shows that 
$\left( \state{\Psi^{\prime}}, \mathcal{W(H)}^{\prime} \right)$ is non-contextual (i.e., $\mathfrak{w}^{\state{\Psi ^{\prime}}}$ has sections). However, $\widehat{\mathcal{W(H)}^{\prime}}$ is non-Boolean, with non-Booleanness $\mathfrak{q} \Gamma \Omega ^{\prime} = 3/5$. Referring to definition \ref{def:l2} and its realisation in the Anders-Browne example (discussion subsequent to \eqref{MermEq}), that $Y_1$ is missing in \eqref{eq:1m} constrains the possible computational inputs and therefore the computation as a whole.  $q_1$ is fixed at $q_1=0$, and hence $i_1 = 0$ and the only computable
function is  $o(0,0)= 0$ and $o(0,1)= 1$, i.e., a partial function on $\mathbb{Z}_2 \times \mathbb{Z}_2$, or 
the identity function on one input.

Supposing, again without loss of generality, that $X_2$ is then measured and the outcome $+1$ is registered.
Then the state after this measurement is

\begin{equation}\state{\Psi ^{\prime \prime}} = \state{++} \otimes (\frac{\state{0} + \state{1}}{\sqrt{2}}) \equiv \state{+++}.
\end{equation}
 Therefore, everything is determined after just two measurements, as would be expected 
due to the constraints on the outcomes. This specifies the poset $V_1$, i.e., local observables $\{X_1, X_2, X_3\}$. The topos $\widehat{V_1} \equiv \widehat{\mathcal{W(H)}^{\prime \prime}} \cong \mathsf{Sets}$ is Boolean,  and of course $\mathfrak{q} \Gamma \Omega ^{\prime \prime} =0$. With the constraints on the computation  $i_1 = i_2 = q_1 = q_2 =q_3= 0$,
the only possible computation is $o(0,0) = 0$.

The situation may be summarised as follows. Measurements $0$, $1$ and $2$ gives rise, for the particular choice of measurements in the above discussion, to the sequence of states

\begin{equation}
\left( \state{\Psi}=\frac{1}{\sqrt{2}} \left( \state{000} + \state{111} \right) \right) \mapsto \left( \state{\psi ^{\prime}} = \frac{1}{\sqrt{2}} \state{+}\otimes (\state{00} + \state{11}) \right) \mapsto \left( \state{\Psi ^{\prime \prime}} = \state{+++} \right),
\end{equation}
specifying a sequence of posets
\begin{equation}
\mathcal{W(H)} \mapsto \mathcal{W(H)}^{\prime} \mapsto \mathcal{W(H)}^{\prime \prime},
\end{equation}
specifying a sequence of topoi 
\begin{equation}
\widehat{\mathcal{W(H)}} \mapsto \widehat{\mathcal{W(H)}^{\prime}} \mapsto \widehat{\mathcal{W(H)}^{\prime \prime}},
\end{equation}
specifying a sequence of Heyting algebras
\begin{equation}
\Gamma \Omega \mapsto \Gamma \Omega ^{\prime} \mapsto \Gamma \Omega ^{\prime \prime},
\end{equation}
specifying a sequence of ``non-Booleannesses"
\begin{equation}
\left( \mathfrak{q}\Gamma \Omega = \frac{111}{113} \right) \mapsto \left( \mathfrak{q}\Gamma \Omega ^{\prime} = \frac{3}{5}\right) \mapsto \left( \mathfrak{q} \Gamma \Omega ^{\prime \prime} = 0 \right);
\end{equation}
accompanied by a sequence of ``computability classes"
\begin{equation}
(\text{All functions on } \mathbb{Z}_2 \times \mathbb{Z}_2) \mapsto (\text{identity function on } \mathbb{Z}_2) \mapsto (0 \mapsto 0).
\end{equation}

Not only have entanglement and state dependent contextuality been exhausted by the computation, we may say that
the  pre-computation non-Booleanness  has been ``consumed" during its course. Since the computational power manifestly
decreases at every stage, we put forward the view that \emph{non-Boolean logic is a computational resource}.

Any deterministic, temporally flat $\ell$-2 MBQC specifies analogous sequences to those given above, always resulting in
a final topos which is Boolean; the relevant post-computation poset in
$\ell$-2 MBQC will be given by a single point of the form $\{O_1(q_1),..., O_n(q_n) \}^{\prime \prime}$ (prime denoting  commutant here).  Hence our findings in the Anders and Browne example have a natural
generalisation to the computations considered in this paper. 

\subsection{Internal perspective}

It is also worth observing that the presheaf $\mathfrak{w}^{\state{\Psi}}$ specifies the computation
``internally" to the topos $\widehat{\mathcal{W(H)}}$.
For the context $V_1$
for example, $\mathfrak{w}^{\state{\Psi}}$
gives rise to the family of sections $\{\lambda_{+++}, \lambda_{+--}, \lambda_{-+-}, \lambda_{--+} \}$, each of which
takes value $1$ on $X_1X_2X_3$ which coincides with the computed output. Identical behaviour occurs for the three remaining contexts (on $V_2$, for instance, we find the family of sections 
which evaluate $X_1X_2X_3$ as $-1$). 

This raises the possibility of a fully internal description of $\ell$-2 MBQC, along with an exploitation of the full power and machinery of topos theory proper. Any topos gives rise to a first order intuitionistic predicate logic - the \emph{Mitchell-Benabou language}. The availability of the quantifiers $\forall$ and $\exists$ allows for the expression of a rich class of theorems; though the techniques required for a proof lie beyond the scope of this paper, it is possible to show that the Anders and Browne OR-gate
takes the form of a theorem within the Mitchell-Benabou language. A full investigation of the scope of this observation remains a goal for the future.

\section{Concluding Remarks and Outlook}

This paper offers a footbridge between the thus far separate disciplines of quantum computation and ``quantum logic", whichever flavour of the latter one wishes to consider. We have established that non-Boolean logic is required for universal classical computation in $\ell$-2 MBQC, consituting a new perspective on classical-over-quantum improvement. Furthermore,
the non-Booleanness appears to be a commodity which is consumed through the act of computation. Interestingly, the specific logical rules which allow for the construction and truth evaluation of sentences in a topos have not been explicitly used; therefore 
further examination of the specific structure of the intuitionistic language
in the topoi arising from MBQC, and what the language tells us about computation, seems a worthy goal.

\section*{Acknowledgements}Thanks are due to Chris Heunen and Rebecca Ronke for valuable feedback on an earlier draft of this manuscript. LL, RD and RR acknowledge support from NSERC, CIFAR and IARPA; LL also acknowledges support from the John Templeton Foundation.

%%%%%%%%%% Insert bibliography here %%%%%%%%%%%%%%

\end{document}